\documentclass{nature}
\usepackage{amsmath}
\usepackage{graphicx}
\usepackage{color}
\usepackage[10pt]{moresize}
\usepackage{graphics,graphicx}
\usepackage{amsfonts}
\usepackage{amssymb}
\usepackage{amscd}
\usepackage{amsmath}
\usepackage{enumerate}
\usepackage{epsfig}
\usepackage{subfigure}

\begin{document}

\title{Fusing atomic $W$ states via quantum Zeno dynamics}
\author{Y. Q. Ji$^{1,2}$, X. Q. Shao$^{1,2}$\footnote{E-mail: shaoxq644@nenu.edu.cn}
\& X. X. Yi$^{1,2}$\footnote{E-mail: yixx@nenu.edu.cn}}
\maketitle
\newcommand{\mm}[1]{\Large $\mathbf{#1}$}
\begin{affiliations}
 \item Center for Quantum Sciences and School of Physics, Northeast Normal University, Changchun 130024, China
 \item Center for Advanced Optoelectronic Functional Materials Research, and Key Laboratory for UV Light-Emitting Materials and Technology
of Ministry of Education, Northeast Normal University, Changchun 130024, China
\end{affiliations}

\baselineskip24pt

\maketitle

\begin{abstract}
We propose a scheme for preparation of large-scale entangled $W$ states based on the fusion mechanism via quantum Zeno dynamics.
By sending two atoms belonging to an $n$-atom $W$ state and an $m$-atom $W$ state, respectively, into a vacuum cavity
(or two separate cavities), we may obtain a ($n+m-2$)-atom $W$ state via detecting the two-atom state after
interaction. The present scheme is robust against both spontaneous emission of atoms and decay of cavity,
and the feasibility analysis indicates that it can also be realized in experiment.
\end{abstract}
\clearpage

\section*{Introduction}
Quantum entanglement, as one of the crucial resources, not only plays a key role in fundamental quantum physics~\cite{001},
but also has wide applications in many quantum information and quantum communication tasks, such as quantum teleportation~\cite{003,004,005},
quantum key distribution~\cite{006,007,008}, quantum secret sharing~\cite{009,010,011,012}, quantum secure direct
communication~\cite{013,014,015,016,017,0171,0172} and so on. Furthermore, it is even considered as an important effect in living biological bodies
in recent years, for instance, the entanglement may be related to Avian compass~\cite{0173}, the entanglement and teleportation using living cell
is also possible~\cite{0174}. In addition, many theoretical and experimental efforts for generating entanglement have been one focus of the current
study~\cite{r3,r4,r5,r6,r7,r8,r9,r10,r11,r12,r13}. Among entangled states, bipartite entangled is the simplest one.
With local operations and classical communication (LOCC), we can obtain an arbitrary bipartite state from a bipartite entangled state. However,
a multipartite entangled state cannot be converted into each other with LOCC~\cite{018,019,020}.

$W$ state is a special kind of entangled state due to its highly robust against the qubit loss.
Hence, $W$ state has always been a hot spot in quantum computing and information science~\cite{021,022}. There are many methods for preparation of
$W$ state, such as Xu \emph{et al}. proposed an efficient scheme to generate multi-photon entangled $W$ state from two-qubit EPR pairs by measurements
and follow-up local transformation~\cite{r3}. Kang \emph{et al}. proposed a protocol to generate a W by using multiple Schr\"{o}dinger dynamics~\cite{r12}
and with superconducting quantum interference devices by using dressed states~\cite{r13}. However, it is difficult to create multipartite $W$
states in a realistic situation because the dynamics becomes more complex as the number of particle increases, which leads to be more sensitive to decoherence.
Thus simple and efficient schemes to prepare large-scale multipartite entangled states are of great importance.
In recent works, quantum state fusion and expansion technology have been put forward to realize large-size multipartite entangled states~\cite{023,024,025,026,027,028,029,030,031,032,033,034,035}. One can get a larger entangled state from two or
more qubits entangled states on the condition that one qubit of each entangled state is sent to the fusion operation~\cite{033}.

Recently, Tashima \emph{et al}. experimentally demonstrated a transformation of two Einstein-Podolsky-Rosen photon pairs
into a three-photon $W$ state using LOCC~\cite{026}. Meanwhile, he also proposed a series of methods to expand polarization entangled
$W$ states~\cite{027,028,029}. In 2011, \"{O}zdemir \emph{et al}. used a simple optical fusion gate to get a
$W_{n+m-2}$ state from $W_{n}$ and $W_{m}$~\cite{033}. In the following years, several $W$ states fusion schemes emerged with the
help of complex quantum gate sets~\cite{030,032}. Nevertheless the realization of Fredkin gate and Toffoli gate are not easy in experiment.
Very recently, Han \emph{et al}. proposed two effective fusion schemes for stationary electronic $W$ state and flying photonic $W$ state, respectively, using
the quantum-dot-microcavity coupled system~\cite{035}, but the schemes are too complicated to be realized.
Meanwhile, Zhang \emph{et al}. also prepared a large-size $W$ state network with a fusion mechanism in cavity QED system~\cite{034}. The quantum information was encoded into the ground state and excited state, which made the fidelity sensitive to spontaneous emission of atoms.

In this paper, we present a theoretical scheme for preparing a large-scale $W$ state via quantum Zeno dynamics in cavity QED system. The interactions
between atoms and the cavity mode are far-off-resonant, which makes the proposed schemes more feasible within the current technology. The fusion operation requires only one particle of each multipartite entangled states sent into an vacuum cavity (or two separate cavities). The success rate for preparing a $W_{n+m-2}$ state depends on the detected states of two atoms. The prominent advantage of our scheme is that the quantum information
is encoded into the ground state, so it is robust against spontaneous emission of atom. In addition, the whole procedure works well in the quantum Zeno subspace, thus the cavity decay has no influence on the evolution of the encoded qubit states.

\section*{Results}
\begin{figure}
\centering\scalebox{0.45}{\includegraphics{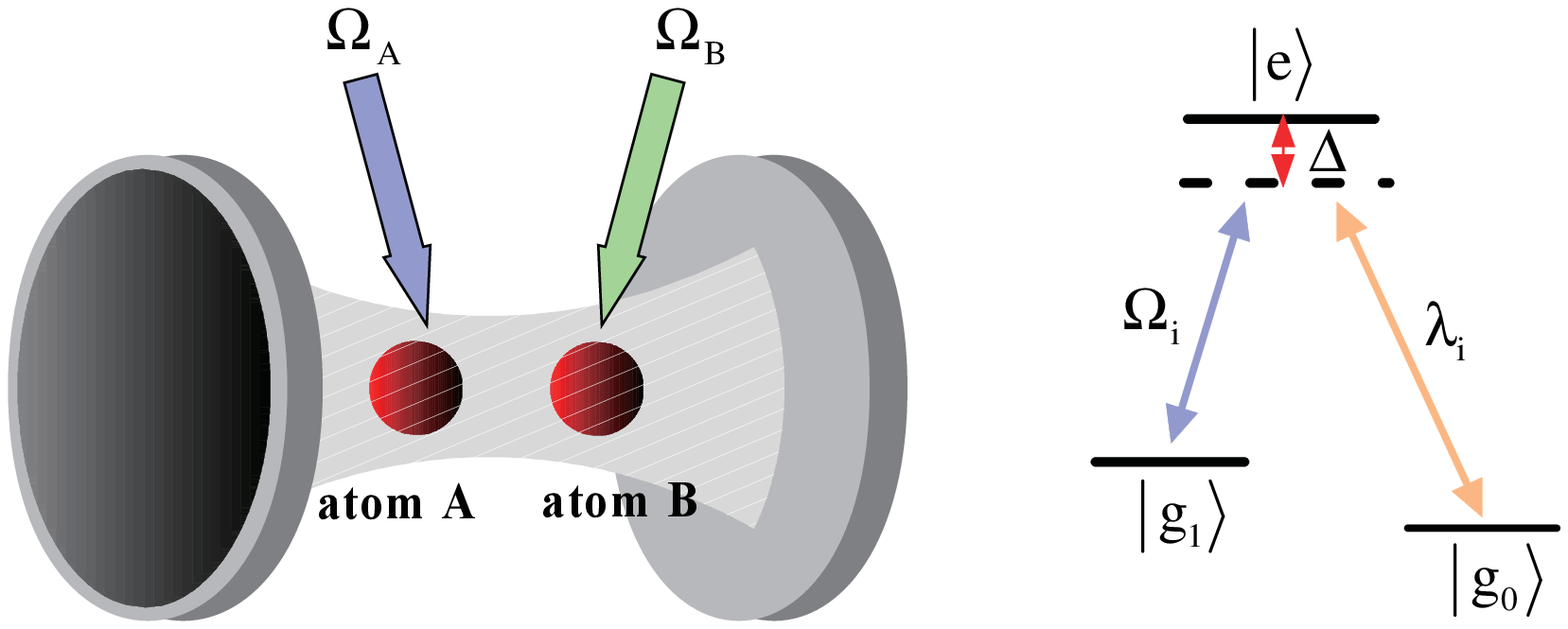}}
\caption{\label{P1}The cavity-atom combined system and the atomic level configuration for the original Hamiltonian.
the transition $|e\rangle\leftrightarrow|g_{1}\rangle$ is driven by classical field with time-dependent Rabi
frequency $\Omega$, the transition $|e\rangle\leftrightarrow|g_{0}\rangle$ is coupled
to the cavity with coupling $\lambda$, and $\Delta$ is detuning parameter.}
\end{figure}

{\bf Fusing atomic $W$ states in a cavity QED system.}
We consider two identical $\Lambda$-type atoms trapped in the cavity, as shown in Fig.~\ref{P1}. Each atom
has an excited state $|e\rangle$ and two ground states $|g_{1}\rangle$ and $|g_{0}\rangle$.
The transition $|e\rangle\leftrightarrow|g_{1}\rangle$ is non-resonantly driven by a classical field with Rabi
frequency $\Omega$ and detuning $\Delta$, the transition $|e\rangle\leftrightarrow|g_{0}\rangle$ is coupled
non-resonantly to the cavity with coupling $\lambda$ and detuning $\Delta$. Under the rotating-wave approximation (RWA),
the interaction Hamiltonian for this system can be written as ($\hbar=1$)
\begin{eqnarray}\label{01}
H_{I}&=&H_{ac}+H_{al}+H_{e}, \cr
H_{ac}&=&\sum_{i=A,B} \lambda_{i}|e\rangle\langle g_{0}|a+\rm{H.c.}, \cr
H_{al}&=&\sum_{i=A,B} \Omega_{i}|e\rangle\langle g_{1}|+\rm{H.c.}, \cr
H_{e}&=&\sum_{i=A,B} \Delta_{i} |e\rangle\langle e|,
\end{eqnarray}
where $a$ denotes annihilation operator of the cavity. For the sake of simplicity, we assume
$\lambda_{A}=\lambda_{B}=\lambda$ and $\Omega_{A}=\Omega_{B}=\Omega$.
Due to the quantum information is encoded in the states $|g_{0}\rangle$ and $|g_{1}\rangle$, there are four possible states
for two atoms, i.e., \{$|g_{0}g_{0}\rangle, |g_{0}g_{1}\rangle, |g_{1}g_{0}\rangle, |g_{1}g_{1}\rangle$\}.

For the initial state of two atoms and cavity is $|g_{0}g_{0}\rangle|0_{c}\rangle$, it is easily to find that the state does
not evolute, because of $H_{I}|g_{0}g_{0}\rangle|0_{c}\rangle=0$.

If the initial state is in $|g_{0}g_{1}\rangle|0_{c}\rangle$ or $|g_{1}g_{0}\rangle|0_{c}\rangle$,
the whole system evolves in a closed subspaces
\{$|g_{0}g_{1}\rangle|0_{c}\rangle, |g_{0}e\rangle|0_{c}\rangle,|g_{0}g_{0}\rangle|1_{c}\rangle,
|eg_{0}\rangle|0_{c}\rangle,|g_{1}g_{0}\rangle|0_{c}\rangle$\}.
Under the Zeno condition $\lambda_{i}\gg\Omega_{i}$, the Hilbert subspace is split into three invariant Zeno subspaces
\begin{eqnarray}\label{02}
Z_{1}&=&\{|g_{0}g_{1}\rangle|0_{c}\rangle,|g_{1}g_{0}\rangle|0_{c}\rangle,|\psi_{1}\rangle\}, \cr
Z_{2}&=&\{|\psi_{2}\rangle\}, \cr
Z_{3}&=&\{|\psi_{3}\rangle\},
\end{eqnarray}
corresponding to the projections $P_{i}=|\alpha\rangle\langle\alpha|$ and $\alpha\in Z_{i}$ ($i=1,2,3$),
where the eigenstates of $H_{ac}$ are
\begin{eqnarray}\label{03}
|\psi_{1}\rangle&=&\frac{1}{\sqrt{2}}(-|g_{0}e\rangle|0_{c}\rangle+|eg_{0}\rangle|0_{c}\rangle), \cr
|\psi_{2}\rangle&=&\frac{1}{2}(|g_{0}e\rangle|0_{c}\rangle-\sqrt{2}|g_{0}g_{0}\rangle|1_{c}\rangle+|eg_{0}\rangle|0_{c}\rangle), \cr
|\psi_{3}\rangle&=&\frac{1}{2}(|g_{0}e\rangle|0_{c}\rangle+\sqrt{2}|g_{0}g_{0}\rangle|1_{c}\rangle+|eg_{0}\rangle|0_{c}\rangle),
\end{eqnarray}
with the corresponding eigenvalues
\begin{eqnarray}\label{04}
\eta_{1}&=&0, \cr
\eta_{2}&=&-\sqrt{2}\lambda, \cr
\eta_{3}&=&\sqrt{2}\lambda.
\end{eqnarray}
Through performing the unitary transformation $U=e^{-i\sum\eta_{i}P_{i}t}$ and neglecting the terms with
high oscillating frequency, we obtain the Hamiltonian
\begin{eqnarray}\label{05}
H_{eff}&=&\frac{\Omega}{\sqrt{2}}(-|g_{0}g_{1}\rangle|0_{c}\rangle+|g_{1}g_{0}\rangle|0_{c}\rangle)\langle\psi_{1}| \cr
&&+\frac{\Omega}{\sqrt{2}}|\psi_{1}\rangle(-\langle0_{c}|\langle g_{1}g_{0}|+\langle0_{c}|\langle g_{0}g_{1}|)+\Delta|\psi_{1}\rangle\langle\psi_{1}|.
\end{eqnarray}
By adiabatically eliminating the state $|\psi_{1}\rangle$ under the condition$\Delta\gg\Omega/\sqrt{2}$,
we then have the final effective Hamiltonian
\begin{eqnarray}\label{06}
H_{fe}&=&-\frac{\Omega^{2}}{2\Delta}(|g_{0}g_{1}\rangle|0_{c}\rangle\langle0_{c}|\langle g_{1}g_{0}|+|g_{1}g_{0}\rangle|0_{c}\rangle\langle0_{c}|\langle g_{0}g_{1}|) \cr
&&+\frac{\Omega^{2}}{2\Delta}(|g_{0}g_{1}\rangle|0_{c}\rangle\langle0_{c}|\langle g_{0}g_{1}|+|g_{1}g_{0}\rangle|0_{c}\rangle\langle0_{c}|\langle g_{1}g_{0}|).
\end{eqnarray}
The first two terms caused by Stark shift can be removed through introducing ancillary classical fields and levels, thus the above Hamiltonian reduce to
\begin{eqnarray}\label{07}
\tilde{H}_{fe}=\frac{\Omega^{2}}{2\Delta}(|g_{0}g_{1}\rangle|0_{c}\rangle\langle0_{c}|\langle g_{0}g_{1}|+|g_{1}g_{0}\rangle|0_{c}\rangle\langle0_{c}|\langle g_{1}g_{0}|).
\end{eqnarray}
Under the application of $\tilde{H}_{fe}$, the dynamical evolution for the initial states $|g_{0}g_{1}\rangle|0_{c}\rangle$ and $|g_{1}g_{0}\rangle|0_{c}\rangle$ become to

\begin{eqnarray}\label{08}
|g_{0}g_{1}\rangle|0_{c}\rangle&\rightarrow& e^{-i\tilde{H}_{fe}t}|g_{0}g_{1}\rangle|0_{c}\rangle \cr
&=&\left[\cos(\frac{\Omega^{2}t}{2\Delta})|g_{0}g_{1}\rangle-i\sin(\frac{\Omega^{2}t}{2\Delta})|g_{1}g_{0}\rangle\right]|0_{c}\rangle, \cr
|g_{1}g_{0}\rangle|0_{c}\rangle&\rightarrow& e^{-i\tilde{H}_{fe}t}|g_{1}g_{0}\rangle|0_{c}\rangle \cr
&=&\left[\cos(\frac{\Omega^{2}t}{2\Delta})|g_{1}g_{0}\rangle-i\sin(\frac{\Omega^{2}t}{2\Delta})|g_{0}g_{1}\rangle\right]|0_{c}\rangle.
\end{eqnarray}
After selecting interaction time $t=\Delta\pi/(2\Omega^{2})$, the above equations leads to
\begin{eqnarray}\label{09}
|g_{0}g_{1}\rangle|0_{c}\rangle\rightarrow\frac{1}{\sqrt{2}}(|g_{0}g_{1}\rangle-i|g_{1}g_{0}\rangle)|0_{c}\rangle,\cr
|g_{1}g_{0}\rangle|0_{c}\rangle\rightarrow\frac{1}{\sqrt{2}}(|g_{1}g_{0}\rangle-i|g_{0}g_{1}\rangle)|0_{c}\rangle.
\end{eqnarray}

If the initial state of atoms is in $|g_{1}g_{1}\rangle|0_{c}\rangle$, the whole system evolves in a closed subspaces
\{$|g_{1}g_{1}\rangle|0_{c}\rangle,|eg_{1}\rangle|0_{c}\rangle,|g_{1}e\rangle|0_{c}\rangle,
|g_{0}g_{1}\rangle|1_{c}\rangle,
|ee\rangle|0_{c}\rangle,|g_{1}g_{0}\rangle|1_{c}\rangle, \\
|g_{0}e\rangle|1_{c}\rangle,
|eg_{0}\rangle|1_{c}\rangle,|g_{0}g_{0}\rangle|2_{c}\rangle$\}.
Similar to the process of Eqs.~(\ref{02})-(\ref{07}), we find that the final effective Hamiltonian $\tilde{H}'_{fe}$
has no effect on the evolution of the state $|g_{1}g_{1}\rangle|0_{c}\rangle$, i.e., $\tilde{H}'_{fe}|g_{1}g_{1}\rangle|0_{c}\rangle=0$.

Due to the above reasons, we can conclude that in the encoded qubit subspace
\{$|g_{0}g_{0}\rangle|0_{c}\rangle$, $|g_{0}g_{1}\rangle|0_{c}\rangle$, $|g_{1}g_{0}\rangle|0_{c}\rangle$, $|g_{1}g_{1}\rangle|0_{c}\rangle$\}, the
temporal evolution takes the form of
\begin{eqnarray}\label{10}
|g_{0}g_{0}\rangle|0_{c}\rangle&\rightarrow&|g_{0}g_{0}\rangle|0_{c}\rangle,\cr
|g_{0}g_{1}\rangle|0_{c}\rangle&\rightarrow&\frac{1}{\sqrt{2}}(|g_{0}g_{1}\rangle-i|g_{1}g_{0}\rangle)|0_{c}\rangle,\cr
|g_{1}g_{0}\rangle|0_{c}\rangle&\rightarrow&\frac{1}{\sqrt{2}}(|g_{1}g_{0}\rangle-i|g_{0}g_{1}\rangle)|0_{c}\rangle,\cr
|g_{1}g_{1}\rangle|0_{c}\rangle&\rightarrow&|g_{1}g_{1}\rangle|0_{c}\rangle.
\end{eqnarray}

\begin{figure}
\centering\scalebox{0.3}{\includegraphics{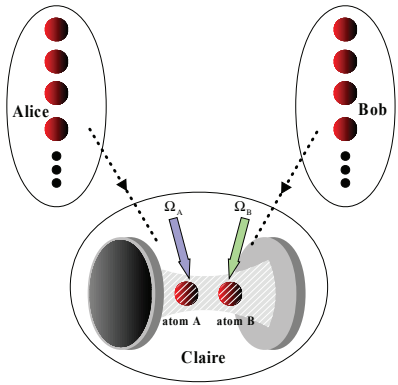}}
\caption{\label{P2} The setup for fusion of two $W$ states. Both Alice and Bob transmit one qubit to Claire, under the condition $t=\Delta\pi/(2\Omega^{2})$,
Claire detects the state of two atoms and informs them if the task is successful.}
\end{figure}

Now, we introduce how to implement a ($m+n-2$) qubits atomic $W$ state fusion scheme from an $m$-qubits $W$ state and an $n$-qubits $W$
state based on quantum Zeno dynamics. As shown in Fig.~\ref{P2}, there are two parties, Alice and Bob, decide to merge their small-scale $|W_{n}\rangle_{A}$ and $|W_{m}\rangle_{B}$ into a larger-scale entangled $W$ state with the help of a third party Claire. In order to do this, each person
transmits one qubit to Claire who received two qubits with quantum Zeno dynamics to merge and informs them when the task is successful.
The atomic entangled $W$ states of Alice and Bob are
\begin{small}
\begin{eqnarray}\label{11}
|W_{n}\rangle_{A}&=&\frac{1}{\sqrt{n}}\left(|(n-1)_{g_{0}}\rangle_{a}|1_{g_{1}}\rangle_{1}+\sqrt{n-1}|W_{n-1}\rangle_{a}|1_{g_{0}}\rangle_{1}\right),\cr
|W_{m}\rangle_{B}&=&\frac{1}{\sqrt{m}}\left(|(m-1)_{g_{0}}\rangle_{b}|1_{g_{1}}\rangle_{2}+\sqrt{m-1}|W_{m-1}\rangle_{b}|1_{g_{0}}\rangle_{2}\right).
\end{eqnarray}
\end{small}
To start the fusion process, the two atoms (atom 1 and atom 2) will be sent into the cavity. So the initial state of the whole system is
\begin{eqnarray}\label{12}
|\phi_{0}\rangle=|W_{n}\rangle_{A}\otimes|W_{m}\rangle_{B}\otimes|0_{c}\rangle
\end{eqnarray}
According the result in Eq.~(\ref{10}), the interaction between the cavity mode and the two atoms will change the initial states into the following state
\begin{eqnarray}\label{13}
|\phi_{1}\rangle&=&\frac{1}{\sqrt{mn}}|(n-1)_{g_{0}}\rangle|(m-1)_{g_{0}}\rangle\otimes|g_{1}\rangle|g_{1}\rangle|0_{c}\rangle \cr
&&+\frac{\sqrt{m-1}}{\sqrt{mn}}|(n-1)_{g_{0}}\rangle|W_{m-1}\rangle\otimes\frac{1}{\sqrt{2}}(|g_{1}g_{0}\rangle-i|g_{0}g_{1}\rangle)|0_{c}\rangle\cr
&&+\frac{\sqrt{n-1}}{\sqrt{mn}}|W_{n-1}\rangle|(m-1)_{g_{0}}\rangle\otimes\frac{1}{\sqrt{2}}(|g_{0}g_{1}\rangle-i|g_{1}g_{0}\rangle)|0_{c}\rangle\cr
&&+\frac{\sqrt{(m-1)(n-1)}}{\sqrt{mn}}|W_{n-1}\rangle_{a}|W_{m-1}\rangle_{b}\otimes|g_{0}g_{0}|0_{c}\rangle.
\end{eqnarray}
Then the two atoms will be detected. The detection result $|g_{0}g_{0}\rangle$ means the failure of the fusion process, the failure probability of $P_{f}=1/mn$. The detection result $|g_{1}g_{1}\rangle$, implies that each of the initial $W$ states has lost one atom, and we will have two
separate $W$ states with a smaller number of qubits, $|W_{n-1}\rangle_{A}$ and $|W_{m-1}\rangle_{B}$, with probability $P_{r}=(n-1)(m-1)/mn$. These shortened $W$ states can be recycled using the same fusion mechanism later.

If the detection result is $|g_{1}g_{0}\rangle$, the remaining atoms are in the following states
\begin{eqnarray}\label{14}
|\phi_{1}\rangle&=&\frac{1}{\sqrt{2mn}}\sqrt{m-1}|(n-1)_{g_{0}}\rangle|W_{m-1}\rangle \cr
&&-\frac{1}{\sqrt{2mn}}\sqrt{n-1}i|W_{n-1}\rangle|(m-1)_{g_{0}}\rangle
\end{eqnarray}
After Alice performs the one-qubit phase gate on all the atoms that she has, i.e., \{$|g_{0}\rangle\rightarrow |g_{0}\rangle,|g_{1}\rangle\rightarrow i|g_{1}\rangle$\}, the states in Eqs.~(\ref{14}) will become
\begin{small}
\begin{eqnarray}\label{15}
|\phi'_{1}\rangle&=&\frac{1}{\sqrt{2mn}}\left(\sqrt{m-1}|(n-1)_{g_{0}}\rangle|W_{m-1}\rangle+\sqrt{n-1}|W_{n-1}\rangle|(m-1)_{g_{0}}\rangle\right)\cr
&=&\frac{\sqrt{n+m-2}}{\sqrt{2mn}}|W_{n+m-2}\rangle,
\end{eqnarray}
\end{small}
where we have used $\sqrt{k}|W_{k}\rangle=\sqrt{i}|W_{i}\rangle|(k-i)_{g_{0}}\rangle+\sqrt{i-1}|i_{g_{0}}\rangle|W_{k-i}\rangle$.
Obviously, $|\phi'_{1}\rangle$ is a atomic $W$ state, i.e., $|W_{n+m-2}\rangle$, and the probability obtaining the
$|\phi'_{1}\rangle$ state is $(n+m-2)/(2mn)$.

If the detection result is $|g_{0}g_{1}\rangle$, the systemic state becomes
\begin{small}
\begin{eqnarray}\label{16}
|\phi_{2}\rangle=&-&\frac{1}{\sqrt{2mn}}i\sqrt{m-1}|(n-1)_{g_{0}}\rangle|W_{m-1}\rangle \cr
&+&\frac{1}{\sqrt{2mn}}\sqrt{n-1}|W_{n-1}\rangle|(m-1)_{g_{0}}\rangle
\end{eqnarray}
\end{small}
After Bob performs the one-qubit phase gate on his atoms, the states in Eqs.~(\ref{16}) will become Eqs.~(\ref{15}), and the corresponding
probability obtained is $(n+m-2)/(2mn)$. Thus the total success probability for the fusion process is
\begin{eqnarray}\label{17}
P_{n+m-2}=\frac{n+m-2}{mn}
\end{eqnarray}

\begin{figure}
\centering\scalebox{0.4}{\includegraphics{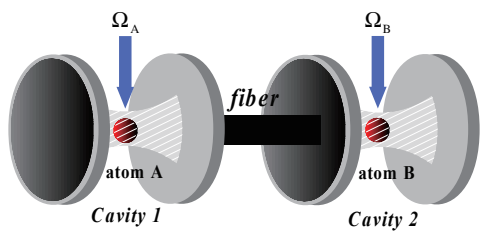}}
\caption{\label{P3} Schematic illustration for Fusing atomic $W$ states in two separate cavities.}
\end{figure}

{\bf Fusing atomic $W$ states in two separate cavities connected by an optical fiber.}
Due to the atoms are trapped in a single cavity, it is hard to control the quantum state. Hence, the other scheme
is proposed for the atoms trapped in different cavities connected by optical fibers. In this section, we will introduce the fusion scheme
of atomic $W$ states in two separate cavities. As shown in Fig.~\ref{P3}, the two atoms, whose level configurations are the same as that
in Fig.~\ref{P1}, are trapped in two cavities connected by a fiber.

In the short fiber limit $L\tau/(2\pi c)\ll1$~\cite{0381,0382}, where $L$ denotes the fiber length, $c$ denotes the
speed of light and $\tau$ denotes the decay of the cavity field into a continuum of fiber mode, only one resonant fiber
mode interacts with the cavity mode. The Hamiltonian
for the cavity-atom-fiber combined system is
\begin{eqnarray}\label{A01}
H'_{I}&=&H'_{ac}+H'_{al}+H'_{e}, \cr
H_{ac}&=&\sum_{i=A,B} \lambda_{i}|e\rangle\langle g_{0}|a_{i}+vb^{\dag}(a_{A}+a_{B})+\rm{H.c.}, \cr
H_{al}&=&\sum_{i=A,B} \Omega_{i}|e\rangle\langle g_{1}|+\rm{H.c.}, \cr
H_{e}&=&\sum_{i=A,B} \Delta_{i} |e\rangle\langle e|,
\end{eqnarray}
where $b^{\dag}$ and $b$ are the creation and annihilation operators for the fiber mode, respectively. $v$ is the coupling strength between
the fiber and the cavities. The same as before, we assume $\lambda_{A}=\lambda_{B}=\lambda$ and $\Omega_{A}=\Omega_{B}=\Omega$.

For the initial state is $|g_{0}g_{0}\rangle|0_{c_{1}}\rangle|0_{c_{2}}\rangle|0_{f}\rangle$, it is easily to find that the state does
not evolute, because of $H'_{I}|g_{0}g_{0}\rangle|0_{c_{1}}\rangle|0_{c_{2}}\rangle|0_{f}\rangle=0$.

If the initial state is in $|g_{0}g_{1}\rangle|0_{c_{1}}\rangle|0_{c_{2}}\rangle|0_{f}\rangle$ or $|g_{1}g_{0}\rangle|0_{c_{1}}\rangle|0_{c_{2}}\rangle|0_{f}\rangle$,
the whole system evolves in a closed subspaces
\begin{eqnarray}\label{A02}
|\varphi_{1}\rangle&=&|g_{1}g_{0}\rangle|0_{c_{1}}\rangle|0_{c_{2}}\rangle|0_{f}\rangle,~|\varphi_{2}\rangle=|eg_{0}\rangle|0_{c_{1}}\rangle|0_{c_{2}}\rangle|0_{f}\rangle,\cr
|\varphi_{3}\rangle&=&|g_{0}g_{0}\rangle|1_{c_{1}}\rangle|0_{c_{2}}\rangle|0_{f}\rangle,~|\varphi_{4}\rangle=|g_{0}g_{0}\rangle|0_{c_{1}}\rangle|0_{c_{2}}\rangle|1_{f}\rangle,\cr
|\varphi_{5}\rangle&=&|g_{0}g_{0}\rangle|0_{c_{1}}\rangle|1_{c_{2}}\rangle|0_{f}\rangle,~|\varphi_{6}\rangle=|g_{0}e\rangle|0_{c_{1}}\rangle|0_{c_{2}}\rangle|0_{f}\rangle,\cr
|\varphi_{7}\rangle&=&|g_{0}g_{1}\rangle|0_{c_{1}}\rangle|0_{c_{2}}\rangle|0_{f}\rangle.
\end{eqnarray}
Under the Zeno condition $\lambda_{i}\gg\Omega_{i}$, the Hilbert subspace is split into five invariant Zeno subspaces
\begin{eqnarray}\label{A03}
Z_{1}&=&\{|\varphi_{1}\rangle,|\varphi_{7}\rangle,|\Phi_{1}\rangle\}, Z_{2}=\{|\Phi_{2}\rangle,\}, \cr
Z_{3}&=&\{|\Phi_{3}\rangle\}, Z_{3}=\{|\Phi_{4}\rangle\}, Z_{5}=\{|\Phi_{5}\rangle\},
\end{eqnarray}
where the eigenstates of $H'_{ac}$ are
\begin{eqnarray}\label{A04}
|\Phi_{1}\rangle&=&N_{1}(|\varphi_{2}\rangle-\alpha|\varphi_{4}\rangle+|\varphi_{6}\rangle),\cr
|\Phi_{2}\rangle&=&N_{2}(-|\varphi_{2}\rangle+|\varphi_{3}\rangle-|\varphi_{5}\rangle+|\varphi_{6}\rangle),\cr
|\Phi_{3}\rangle&=&N_{3}(-|\varphi_{2}\rangle-|\varphi_{3}\rangle+|\varphi_{5}\rangle+|\varphi_{6}\rangle),\cr
|\Phi_{4}\rangle&=&N_{4}(|\varphi_{2}\rangle-\beta|\varphi_{3}\rangle+\gamma|\varphi_{4}\rangle-\beta|\varphi_{5}\rangle+|\varphi_{6}\rangle),\cr
|\Phi_{5}\rangle&=&N_{5}(|\varphi_{2}\rangle+\beta|\varphi_{3}\rangle+\gamma|\varphi_{4}\rangle+\beta|\varphi_{5}\rangle+|\varphi_{6}\rangle),
\end{eqnarray}
with the corresponding eigenvalues
\begin{eqnarray}\label{A05}
\eta_{1}=0,~~\eta_{2}=-\lambda,~~\eta_{3}=\lambda,\cr
\eta_{4}=-\sqrt{2v^{2}+\lambda^{2}},~~\eta_{5}&=&\sqrt{2v^{2}+\lambda^{2}},
\end{eqnarray}
where the parameters are
\begin{eqnarray}\label{A06}
\alpha=\frac{\lambda}{v},~~
\beta=\frac{\sqrt{2v^{2}+\lambda^{2}}}{\lambda},~~
\gamma=\frac{2v}{\lambda},
\end{eqnarray}
in addition, $N_{i}$ is the normalization factor of the eigenstate $|\Phi_{i}\rangle$ (i=1,2,...,5).
Through performing the unitary transformation $U=e^{-i\sum\eta_{i}P_{i}t}$ and neglecting the terms with
high oscillating frequency with setting the Zeno condition, we obtain the Hamiltonian
\begin{eqnarray}\label{A07}
H'_{eff}=N_{1}\Omega(|\varphi_{1}\rangle\langle\Phi_{1}|+|\varphi_{7}\rangle\langle\Phi_{1}|)+H.c.+2\Delta N_{1}^{2}|\Phi_{1}\rangle
\end{eqnarray}
By adiabatically eliminating the state $|\Phi_{1}\rangle$, we obtain the final effective Hamiltonian
\begin{eqnarray}\label{A08}
H'_{fe}=-\frac{\Omega^{2}}{2\Delta}(|\varphi_{1}\rangle\langle\varphi_{1}|+|\varphi_{7}\rangle\langle\varphi_{7}|
+|\varphi_{1}\rangle\langle\varphi_{7}|+|\varphi_{7}\rangle\langle\varphi_{1}|)
\end{eqnarray}
After removed the first two terms ($|\varphi_{1}\rangle\langle\varphi_{1}|,|\varphi_{7}\rangle\langle\varphi_{7}|$) caused by Stark shift,
the above Hamiltonian becomes
\begin{eqnarray}\label{A09}
\tilde{H}'_{fe}&=&-\frac{\Omega^{2}}{2\Delta}(|\varphi_{1}\rangle\langle\varphi_{7}|+|\varphi_{7}\rangle\langle\varphi_{1}|).
\end{eqnarray}
Under the condition $t=\Delta\pi/(2\Omega^{2})$, it leads to
\begin{small}
\begin{eqnarray}\label{A10}
|g_{0}g_{1}\rangle|0_{c_{1}}\rangle|0_{c_{2}}\rangle|0_{f}\rangle
&\rightarrow&\frac{1}{\sqrt{2}}(|g_{0}g_{1}\rangle+i|g_{1}g_{0}\rangle)|0_{c_{1}}\rangle|0_{c_{2}}\rangle|0_{f}\rangle,\cr
|g_{1}g_{0}\rangle|0_{c_{1}}\rangle|0_{c_{2}}\rangle|0_{f}\rangle
&\rightarrow&\frac{1}{\sqrt{2}}(|g_{1}g_{0}\rangle+i|g_{0}g_{1}\rangle)|0_{c_{1}}\rangle|0_{c_{2}}\rangle|0_{f}\rangle.
\end{eqnarray}
\end{small}
If the initial state of atoms is in  $|g_{1}g_{1}\rangle|0_{c_{1}}\rangle|0_{c_{2}}\rangle|0_{f}\rangle$,
similar to the process of Eqs.~(\ref{A02})-(\ref{A08}), we find that the final effective Hamiltonian
has no effect on the evolution of the state $|g_{1}g_{1}\rangle|0_{c_{1}}\rangle|0_{c_{2}}\rangle|0_{f}\rangle$.

According to the results of the above, the temporal evolution takes the form of
\begin{eqnarray}\label{A11}
|g_{0}g_{0}\rangle|0_{c_{1}}\rangle|0_{c_{2}}\rangle|0_{f}\rangle&\rightarrow&|g_{0}g_{0}\rangle|0_{c_{1}}\rangle|0_{c_{2}}\rangle|0_{f}\rangle,\cr
|g_{0}g_{1}\rangle|0_{c_{1}}\rangle|0_{c_{2}}\rangle|0_{f}\rangle&\rightarrow&\frac{1}{\sqrt{2}}(|g_{0}g_{1}\rangle+i|g_{1}g_{0}\rangle)
|0_{c_{1}}\rangle|0_{c_{2}}\rangle|0_{f}\rangle,\cr
|g_{1}g_{0}\rangle|0_{c_{1}}\rangle|0_{c_{2}}\rangle|0_{f}\rangle&\rightarrow&\frac{1}{\sqrt{2}}(|g_{1}g_{0}\rangle+i|g_{0}g_{1}\rangle)
|0_{c_{1}}\rangle|0_{c_{2}}\rangle|0_{f}\rangle,\cr
|g_{1}g_{1}\rangle|0_{c_{1}}\rangle|0_{c_{2}}\rangle|0_{f}\rangle&\rightarrow&|g_{1}g_{1}\rangle|0_{c_{1}}\rangle|0_{c_{2}}\rangle|0_{f}\rangle.
\end{eqnarray}

Now, we use a similar method to fusing atomic $W$ states in two separate cavities. For $m$ qubits $W$ state and $n$ qubits $W$ as shown in Eq.~(\ref{11}),
Alice and Bob transmits one qubit to Claire. The two atoms will be sent into two cavities. According the result in Eq.~(\ref{A11}), two atoms will
evolve to the following state
\begin{small}
\begin{eqnarray}\label{A12}
|\phi_{1}\rangle&=&\frac{1}{\sqrt{mn}}|(n-1)_{g_{0}}\rangle|(m-1)_{g_{0}}\rangle\otimes|g_{1}\rangle|g_{1}\rangle|0_{c_{1}}\rangle|0_{c_{2}}\rangle|0_{f}\rangle \cr
&&+\frac{\sqrt{m-1}}{\sqrt{2mn}}|(n-1)_{g_{0}}\rangle|W_{m-1}\rangle\otimes(|g_{1}g_{0}\rangle+i|g_{0}g_{1}\rangle)
|0_{c_{1}}\rangle|0_{c_{2}}\rangle|0_{f}\rangle\cr
&&+\frac{\sqrt{n-1}}{\sqrt{2mn}}|W_{n-1}\rangle|(m-1)_{g_{0}}\rangle\otimes(|g_{0}g_{1}\rangle+i|g_{1}g_{0}\rangle)
|0_{c_{1}}\rangle|0_{c_{2}}\rangle|0_{f}\rangle\cr
&&+\frac{\sqrt{(m-1)(n-1)}}{\sqrt{mn}}|W_{n-1}\rangle_{a}|W_{m-1}\rangle_{b}\otimes|g_{0}g_{0}\rangle|0_{c_{1}}\rangle|0_{c_{2}}\rangle|0_{f}\rangle
\end{eqnarray}
\end{small}
After the two atoms are detected, the detection result $|g_{0}g_{0}\rangle$ means the failure of the fusion process, and $|g_{1}g_{1}\rangle$ implies we obtain two separate $W$ states with a smaller number of qubits. If the detection result is $|g_{1}g_{0}\rangle$, Bob need to perform the one-qubit phase gate on all the atoms that he has. If the detection result is $|g_{0}g_{1}\rangle$, then Alice performs the one-qubit phase gate on her atoms. Note that, who need to perform the one-qubit phase gate is different from the previous but just the opposite with before. In this process we ignore the global phase. The total success probability is also $(n+m-2)/(mn)$.

\section*{Discussion}

\begin{figure}
\centering\scalebox{0.75}{\includegraphics{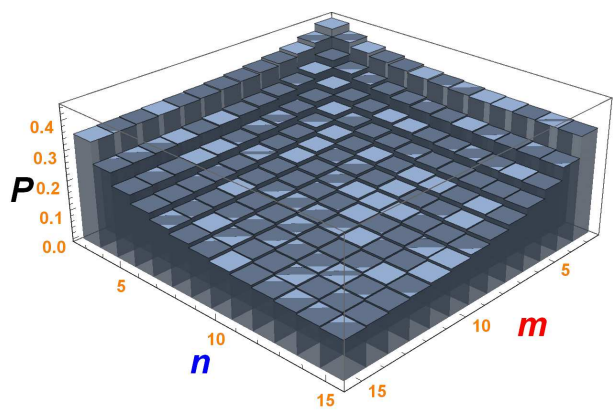}}
\caption{\label{PX}The total success probability of $W$ state fusion scheme varies with $m$ and $n$.}
\end{figure}

For the previous two schemes, both of the total success probability are $(n+m-2)/(mn)$, we plot the success probability varies with $m$ and $n$
in Fig.~\ref{PX}. One can see that the success probability decreases with increasing of $m$ and $n$.
In addition, we know that the Zeno condition $\lambda_{i}\gg\Omega_{i}$ is the precondition for the scheme implementation. Next, we
discuss how to properly choose parameters to satisfy the Zeno condition. Now we give an assessment of the performance
when the fusion scheme is put into practice. In the present model, the dissipation channels include NV centre spontaneous
decay $\gamma$ and photon leakage out of the cavity $\kappa$. When these decoherence effects are taken into account and
under the assumptions that the decay channels are independent, the master equation of the whole system can be expressed
by the Lindblad form~\cite{039,040}
\begin{eqnarray}\label{F1}
\dot{\rho}=&-&i[H,\rho]-\frac{\kappa}{2}\left(a^{\dag}a\rho-2a\rho a^{\dag}+\rho a^{\dag}a\right) \cr
&-&\frac{1}{2}\sum_{k=1}^{4}\left[\hat{\mathcal{L}}_{k}^{\dag}\hat{\mathcal{L}_{k}}\rho-2\hat{\mathcal{L}_{k}}\rho\hat{\mathcal{L}_{k}^{\dag}}
+\rho\hat{\mathcal{L}_{k}^{\dag}}\hat{\mathcal{L}_{k}}\right],
\end{eqnarray}
where $\kappa$ denotes the decay rate of the cavity, $\hat{\mathcal{L}_{1}}=\sqrt{\gamma/2}|g_{0}\rangle_{1}\langle e|$, $\hat{\mathcal{L}_{2}}=\sqrt{\gamma/2}|g_{1}\rangle_{1}\langle e|$, $\hat{\mathcal{L}_{3}}=\sqrt{\gamma/2}|g_{0}\rangle_{2}\langle e|$ and $\hat{\mathcal{L}_{4}}=\sqrt{\gamma/2}|g_{1}\rangle_{2}\langle e|$ are Lindblad operators that describe the dissipative processes.

\begin{figure}
\centering\scalebox{0.5}{\includegraphics{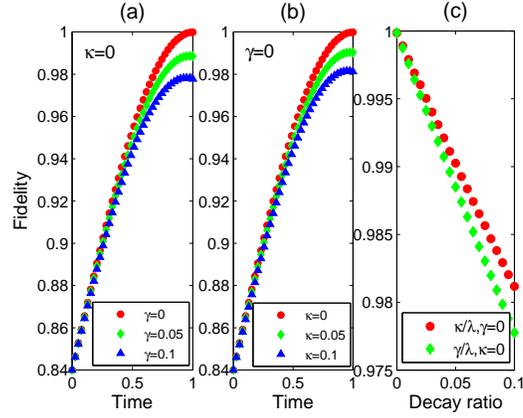}}
\caption{\label{P4}The fidelity of $W$ state fusion scheme for the two atoms in one cavity with $\lambda=1$, $\Omega=0.01\lambda$, $\Delta = 0.8\lambda$.
(a)Fidelity of the fusion varies with $t$ when $\gamma=0, 0.05\lambda, 0.1\lambda$, respectively.
(b)Fidelity of the fusion varies with $t$ when $\kappa=0, 0.05\lambda, 0.1\lambda$, respectively.
(c)Fidelity of the fusion varies with decay ratio. Red circle is the fidelity varies with $\kappa/\lambda$ when $\gamma=0$.
Green rhombus is the fidelity varies with $\gamma/\lambda$ when $\kappa=0$.}
\end{figure}

We use the Eq.~(\ref{13}) act as the ideal final state to check the performance of our scheme, where $m=n=5$.
The fidelity is defined as $\langle\psi_{ideal}|\hat{\rho}(t)|\psi_{ideal}\rangle$.
Fig.~\ref{P4} shows that the relationship between the fidelity and the parameters $t$, $\kappa$ and $\gamma$, and find that the fusion can be finished at time $\frac{\Delta\pi}{2\Omega^{2}}$, and it is immune to both the cavity decay and the spontaneous emission,
since for a large decay condition $\kappa=\gamma=0.1\lambda$, the fidelity remains 96\%. This is because that in the Zeno subspace,
the state of the cavity is always in the vacuum state, hence, the cavity decay terms have no influence on the evolution of the encoded qubit states.
The further large detuning condition excludes the excited states, so this process is also robust against the decoherence induced by
spontaneous emission. In a real experiment, the $\Lambda$ configuration can be found in the cesium atoms which is trapped in a small optical cavity
in the strong-coupling regime~\cite{041,042} can be used in this scheme. Furthermore, a set of cavity quantum
electrodynamics parameters $(\lambda,\gamma,\kappa)/2дл=(750, 2.62, 3.5)$ MHz in strong-coupling regime~\cite{043,044,045},
we can achieve the fusion with a fidelity 99.8\%. Also we can consider the other system, i.e., N-V centre with two unpaired electrons located at the vacancy
and the corresponding experimental parameters $g=2\pi\times2.25$ GHz, $\gamma=2\pi\times0.013$ GHz and $\kappa=2\pi\times0.16$ GHz, we can also achieve the fusion with a fidelity 99.5\% when $\Omega=0.001\lambda$.

\begin{figure}
\centering\scalebox{0.6}{\includegraphics{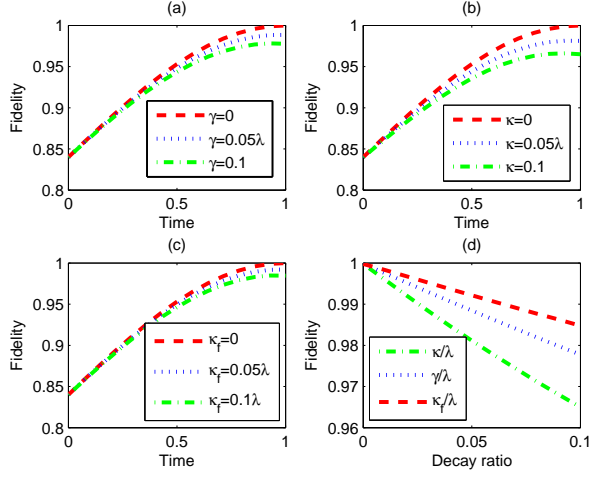}}
\caption{\label{P5}The fidelity of $W$ state fusion scheme for the two atoms in two separate cavities with $\lambda=1$, $v=\lambda$,
$\Omega=0.01\lambda$, $\Delta = 0.8\lambda$. (a)Fidelity of the fusion varies with $t$ when $\gamma=0, 0.05\lambda, 0.1\lambda$, respectively.
(b)Fidelity of the fusion varies with $t$ when $\kappa=0, 0.05\lambda, 0.1\lambda$, respectively.
(c)Fidelity of the fusion varies with $t$ when $\kappa_{f}=0, 0.05\lambda, 0.1\lambda$, respectively.
(d)Fidelity of the fusion varies with decay ratio. Green dot dashed line is the fidelity varies with $\kappa/\lambda$ when $\gamma=0,\kappa_{f}=0$.
Blue dot line is the fidelity varies with $\gamma/\lambda$ when $\kappa=0,\kappa_{f}=0$. Red dashed line is the fidelity varies with
$\kappa_{f}/\lambda$ when $\kappa=0,\gamma=0$.}
\end{figure}

For the cavity-atom-fiber system, he fiber loss at 852 nm wavelength is only about 2.2 dB/Km~\cite{046}, in this case, the fiber decay rate is only 0.152 MHz.
This means that the fiber decay can actually be neglected in a real experiment. In Fig.~\ref{P5}, we use the Eq.~(\ref{A12}) act as the ideal final state to check the performance of our scheme and plot the fidelity for fusing $W$ states and shows that the fidelity versus $t$, $\kappa$, $\kappa_{f}$ and $\gamma$, where $\kappa_{f}$ is the decay of fiber. The fidelity also can reach 99.7\%. Even though we choose to another system (the N-V centre located at the vacancy),
the fidelity still can achieve 99.4\%.

In summary, we have proposed a scheme to fuse a large-scale entangled $W$ states using quantum Zeno dynamics.
The advantages of our scheme is the quantum information is encoded in the ground state and against for spontaneous emission of atom
and cavity decay. Final numerical simulation based on one group of experiment parameters shows that our scheme could be feasible
under current technology and have a high fidelity.\\

\section*{Method}

The key step of our fusion schemes is using quantum Zeno dynamics induced by continuous coupling~\cite{0361,0371}.
The quantum Zeno dynamics was named by Facchi and Pascazio in 2002~\cite{0361}. It is derived from the quantum Zeno effect which describes an
especially phenomenon that transitions between quantum states can be hindered by frequent measurement. In fact, the system can actually evolve away
from its initial state and remain in the Zeno subspace defined by the measurement when frequently projected onto a multidimensional
subspace. In accordance with von Neumann's projection postulate, the quantum Zeno dynamics can be obtained with continuous coupling between
the system and an external system. In general, we assume that a dynamical evolution process is governed by the Hamiltonian
$H_{K}=H_{obs}+KH_{meas}$, where $H_{obs}$ is the Hamiltonian of the subsystem to be investigated, $H_{meas}$ is an additional interaction
Hamiltonian that performs the measurement, and $K$ is the corresponding coupling constant. Consider the time evolution operator
\begin{eqnarray}\label{M1}
U_{K}(t)=\exp(-iH_{K}t),
\end{eqnarray}
For a strong coupling limit $K\rightarrow \infty$, the dominating contribution is $\exp(-iKH_{meas}t)$. Thus we consider limiting evolution operator
\begin{eqnarray}\label{M2}
\mathcal{U}(t)=\lim_{K\rightarrow \infty}\exp(iKH_{meas}t)U_{K}(t),
\end{eqnarray}
which can be shown to have the form~\cite{0361}
\begin{eqnarray}\label{M3}
\mathcal{U}(t)=\exp(-iKH_{Z}t),
\end{eqnarray}
where $H_{Z}=\sum_{n}P_{n}H_{obs}P_{n}$ is the Zeno Hamiltonian and $P_{n}$ is the eigenprojection of the $H_{meas}$ belonging to the
eigenvalue $\lambda_{n}$
\begin{eqnarray}\label{M4}
H_{meas}=\sum_{n}\lambda_{n}P_{n},
\end{eqnarray}
Therefor, the limiting evolution operator is
\begin{eqnarray}\label{M5}
U_{K}(t)\sim\exp(-iH_{meas}t)\mathcal{U}(t)=\exp\left(-i\sum_{n}Kt\lambda_{n}P_{n}+P_{n}H_{obs}P_{n}t\right),
\end{eqnarray}
corresponding to an effective Hamiltonian
\begin{eqnarray}\label{M6}
H_{eff}=\sum_{n}K\lambda_{n}P_{n}+P_{n}H_{obs}P_{n}
\end{eqnarray}
If the system is initialized in the dark state with respect to $H_{meas}$, the effective Hamiltonian will be reduced to $H_{Z}$.
This new finding has enlightened many works in quantum information processing tasks~\cite{0372,0373,0374,0375,0376,0377,0378,0379,03791}.

\noindent\textbf{Acknowledgments}\\
This work is supported by National Natural Science
Foundation of China (NSFC) under Grants No. 11534002, No. 61475033
and Fundamental Research Funds for the
Central Universities under Grants No. 2412016KJ004.

\noindent\textbf{Author Contributions}\\
X. Q. Shao and X.X.Yi contributed the idea.
Y. Q. Ji performed the calculations and numerical calculations.
Y. Q. Ji wrote the main manuscript, X. Q. Shao and X.X.Yi checked the calculations and made an improvement of the manuscript.
All authors contributed to discussion and reviewed the manuscript.

\noindent\textbf{Additional Information}\\
The authors declare no competing financial interests.



\clearpage

\clearpage

\clearpage

\clearpage

\clearpage

\clearpage

\end{document}